\date{\today}
\documentclass[12pt]{article}
\usepackage{url}
\usepackage{graphicx}
\usepackage{float}
\usepackage{sansmath}
\usepackage{xcolor}
\usepackage[affil-it]{authblk}
\begin{document} \sloppy
\renewcommand\Authfont{\fontsize{12}{10}\selectfont}
\renewcommand\Affilfont{\fontsize{6}{0}\itshape}
\title{\huge The role of low temperature waste heat recovery in achieving 2050 goals: a policy positioning paper}
\author[1]{Edward Wheatcroft}
\author[1]{Henry Wynn}
\author[2]{Kristina Lygnerud}
\author[3]{Giorgio Bonvicini}
\affil[1]{\tiny Centre for the Analysis of Time Series, London School of Economics, London, WC2A 2AE, UK}
\affil[2]{\tiny IVL Swedish Environmental Research Institute, G{\"o}teborg, 41133, Sweden}
\affil[3]{\tiny RINA Consulting S.p.A. Genova, Italy}
\maketitle
\noindent
\normalsize

\begin{abstract}
Urban waste heat recovery, in which low temperature heat from urban sources is recovered for use in a district heat network, has a great deal of potential in helping to achieve 2050 climate goals.  For example, heat from data centres, metro systems, public sector buildings and waste water treatment plants could be used to supply ten percent of Europe's heat demand.  Despite this, at present, urban waste heat recovery is not widespread and is an immature technology. To help achieve greater uptake, three policy recommendations are made. First, policy raising awareness of waste heat recovery and creating a legal framework is suggested.  Second, it is recommended that pilot projects are promoted to help demonstrate technical and economic feasibility.  Finally, a pilot credit facility is proposed aimed at bridging the gap between potential investors and heat recovery projects. 
\end{abstract}



\section{Introduction}
Conventional district heating (DH) systems are a combination of heat from recycled sources (combined heat and power generation, waste to energy plants and industrial processes) and renewable sources such as biomass fuels and solar collectors (\cite{werner2017international}). The conventional wisdom is that heat is collected or produced locally and distributed to customers in a citywide distribution network (\cite{lygnerud2018challenges}). The most recent technological development is the usage of local, low temperature heat sources stemming from city infrastructure such as metro systems and sewage and commercial activity from sources such as data centres (\cite{lygnerud2019contracts}). Low temperature heat recovery is referred to as 4th and 5th generation district heating technology where the 5th generation represents lower distribution temperatures than the 4th. The technology shift is, at present, new and far from established. However, utilisation of urban heat sources will become increasingly important in a future with no fossil fuels, lower available volumes of waste to incinerate and demand for alternative uses of biofuels. This presents an opportunity since such sources are numerous and tend to be situated close to urban areas of high heat demand; aspects that reinforce the resilience of the energy system overall. \par
 
The combination of industrial and urban waste heat sources provides great potential for the replacement of fossil fuel based heating in Europe and globally. The global extent of waste heat recovery from industrial sources is not known. However, it is estimated that, in 2014, 331 PJ of industrial heat was recovered, whilst 16 PJ was recovered in Sweden. The potential for industrial heat recovery in district heating systems has been estimated for the UK (\cite{mckenna2010spatial}), Spain (\cite{miro2016methodologies}), Germany (\cite{miro2016methodologies, bruckner2014using}), China (\cite{fang2015key, li2016case}) and the EU as a whole (\cite{connolly2013heat}). The potentials are high but, in all cases, there is a significant risk that the industrial waste heat provider will cease to operate (\cite{werner2017international}). \par

Around 2.7 EJ/year of industrial waste heat is estimated to be available in Europe. Additionally, around 1.2 EJ/year of low temperature heat is available from data centres, metro stations, waste water treatment plants and service sector buildings (\cite{lygnerud2019contracts}). For context, the EU's total demand for domestic heat and hot water is estimated at 10.7 EJ/year (\cite{persson2018accessible}). Despite the great potential for urban waste heat recovery, uptake has, so far, been low. Research has shown that urban waste heat recovery investment is hindered by untested technical solutions, the absence of standardised contracts and the current cost of carbon, making the business model more appropriate for circumstances beyond 2050 (\cite{lygnerud2019contracts}). \par

This paper was written from the perspective of the ReUseHeat project, funded by the European Commission. The project demonstrates four system innovations for recovering urban waste heat and conducts research into stakeholder perspectives, investment risks, requirements of investors and business models. The project is a forerunner in terms of low temperature waste heat recovery and showcases what the district heating sector may look like in 2050. The aim of this paper is to discuss the role of policy in encouraging urban waste heat recovery for the heating and cooling sector in Europe.  \par

\section{Literature review} \label{section:literature_review}
\subsection{Heat market overview}
In the EU, district heating has an overall 13 percent market share of the heating sector.  However, in some markets, there is a longer tradition and the market is more mature.  District heating contributes to over 50 percent of heat demand in Denmark, Sweden, Finland, Estonia, Latvia, Lithuania and Poland (\cite{werner2017international}). In terms of absolute numbers, however, the biggest district heating markets in the EU can be found in Germany and Poland (\cite{ECOHEAT4EU}). In Germany, the prevalence of district heating is higher in eastern states (approximately 30 percent) than in western states (around 9 percent) (\cite{ECOHEAT4EU}). The European district energy landscape can be split into four categories: consolidation countries, refurbishment countries, expansion countries and new developing countries (\cite{aronsson2011existing}). \par

\subsection{EU frameworks that impact district heating}
Frameworks that impact the implementation of district energy in the EU originate both from EU directives and national legislation.  Legal frameworks for district heating differ significantly between regions and countries. Laws explicitly targeting district heating are in place in Denmark, Sweden, Norway, Lithuania and Germany. In other countries, such as Finland, France, the UK and the USA, energy and competition laws dominate and apply to district energy markets (\cite{werner2017international}). \par

In 2016, the European Commission drafted a strategy for district heating and cooling which, among other things, aims to better integrate it with the electricity system. Further, the strategy aims to reduce energy waste in industry by utilising waste heat in district heating systems (\cite{lygnerud2019contracts}). However, the strategy is not currently implemented across the European Union in a structure that resembles the directive. In a study of 14 EU countries (\cite{aronsson2011existing}), directives with an impact on district energy have been identified. These are:

\begin{itemize}
\item European Parliament Directive 2002/91/EC on the energy performance of buildings.
\item European Parliament Directive 2004/8/EC on the promotion of cogeneration, based on useful heat demand in the internal energy market, amending Directive 92/42/EEC.
\item European Parliament Directive 2006/32/EC of the European Parliament and of the Council on energy end-use efficiency and energy services and Energy Efficiency Directive 2012/12/EU.
\item European Parliament Directive 2008/1/EC concerning integrated pollution prevention and control (IPPC).
\item European Parliament Directive 2008/98/EC of the European Parliament on waste and repealing certain Directives.
\item European Parliament Directive 2009/28/EC of the European Parliament on the promotion of the use of energy from renewable sources.
\end{itemize}

Different countries within the EU have different traditions towards regulation. Below, countries adhering to the tradition of explicit heat regulation and those adhering to broader energy and competition regulation are discussed. At the end of the section, lessons that can be learned from these examples are discussed. \par

\subsection{Countries with explicit heat regulations}
Two examples of countries with explicit heat regulations and mature heat markets are now discussed (Denmark and Sweden). Both countries have explicit heat regulation whilst one market is regulated (Denmark) and the other is privatised (Sweden). \par 

\subsubsection{Denmark}
In Denmark, there are several measures in place that support district heating.  These are (i) heat planning regulation, (ii) taxation, (iii) subsidies, (iv) heat price regulation, (v) CHP requirements, (v) a ban on electrical heating and (vi) a law on district cooling. There is a tradition of encouraging efficient use of energy by ensuring there is a market for collective heat supply. This is achieved through heat planning regulation. This planning also encompasses planning for the usage of waste. Taxation is applied on energy as a fuel and on its emissions, promoting a shift from fossil fuels. Before the millennium, there were different, direct subsidies impacting district heating usage (such as subsidies for converting older houses to DH, speeding up the process of planned networks and conversion from coal to gas). The heat price is regulated to reflect the actual cost. The idea is to outweigh the disadvantage to consumers of the natural monopolies that come from district heating ventures. The current electricity act stipulates that CHPs are built with the main ambition of generating electricity. However, heat recovery from CHPs is mandatory. A ban on electric heating in buildings also encourages DH. The last support mechanism for DH is that there is law that allows municipalities to operate commercial district cooling schemes using the same infrastructure as for DH (\cite{aronsson2011existing}). \par

In Denmark, in March 2012, parliament agreed on a 2020 climate strategy and to a long-term target to reach 100 percent renewable energy in the energy system by 2050. In 2018, the Danish government signed a new energy agreement with the support of all sitting parties in parliament reaffirming and strengthening Denmark's climate and energy goals leading up to 2030. The energy agreement contains a wide range of ambitious green initiatives. Companies and consumers will, in the upcoming years, receive cheaper heating through a modernisation of the heating sector, where both the district heating sector and consumers will have a free choice to decide on future investments.  This will result in cheap heating both for companies and consumers (\cite{tian2019large}). Since the Danish government imposes an energy tax on natural gas, solar district heating plants are given the opportunity to compete with natural gas boilers. Consequently, solar district heating plants are commercially viable solutions in Danish district energy systems (\cite{tian2019large}). Taking the interest of low temperature district heating investments into account, it is noted that, in Denmark, the 4DH research centre was active between 2012 and 2018 with contributions from several Danish and international universities, together with many Danish district heating and companies. Researchers within this centre have written a 4GDH definition paper (\cite{lund20144th}) and a 4GDH status paper (\cite{lund2018status}). Many papers have also been published in scientific journals from the annual international 4DH conferences since 2015 (\cite{lund2019perspectives}). Other than this initiative, however, there are no dedicated low temperature district heating incentive in place. \par

\subsubsection{Sweden}
With the support of energy and climate policy, district heating systems have been developed since the 1970's and transformed from a dependence on fossil fuels to biomass. This progression has resulted in the DH sector being almost fossil free (\cite{lygnerud2018challenges}). Sweden, for example, has the largest percentage of industrial heat recovery in its district heating systems in the world \cite{lygnerud2017risk}.  An energy and carbon tax was introduced on fuels used in heat production in 1991. From 2008 onwards, the carbon tax was gradually phased out for those combustion plants covered by the EU Emission Trading Scheme (EU-ETS), in order not to interfere with this policy. Two governmental investment grant schemes (1991-1996 and 1997-2002) played an important role in the construction phase of biomass-based CHPs. A scheme for Tradable Renewable Electricity Certificates (TRC) was introduced in 2003 in order to support electricity from renewable energy sources and peat. This led to additional CHP production and, consequently, a shift from fossil fuels to biomass in CHP plants (\cite{bouchikhi2003escaping,brange2016prosumers}). \par

In 1996, the Swedish heat market was deregulated. Prior to this, the largest challenge to the district heating industry was to improve its production technology to meet the increasing demand (\cite{lygnerud2006value}), but, since 1996, the industry has had to cope with challenges beyond technology. Examples of institutional challenges are a new district heating law (2008), the threat of price regulation (2009) and third party access (2009).  The Energy Efficiency Plan 2011 and the Energy Efficiency Directive 2012/12/EU (also known as EED) promote effective heat recovery systems from electricity and industrial production processes as a way to help reach the EU target. The national proposal for the implementation of the EED in Sweden states that a cost-benefit analysis should be performed to evaluate investments in the use of excess heat in comparison with other thermal supply systems (\cite{energisverige}). The Swedish government bill states that a DH company has no obligation to allow regulated access ``if it can show that there is a risk that it will suffer damage as a result of the access''. This means that Swedish policy currently neither promotes nor advises against using excess heat recovery in DH (\cite{energisverige}). On December 4th, 2019 it was announced that new legislation placing a tax on waste incineration will be introduced from April 1st, 2020. In combination with the new tax, the energy and $C0_{2}$ tax on CHP in will increase from 30 percent to 100 percent and 11 percent to 91 percent respectively. These new taxes will be a challenge to Swedish CHPs. Taking the interests of low temperature district heating investments into account, it is noted that no regulation explicitly supporting such investments exists in Sweden (\cite{reuseheat_2_1}).

\subsection{Countries with energy and competition law regulation}
Germany, France and the UK are examples of countries with energy and competition law regulation.  Out of these, Germany is a mature market, France has had a district heating market for a long time which is currently experiencing significant expansion and the UK is a new district heating market. Each of these countries are now discussed.

\subsubsection{Germany}
The district-heating sector in Germany has a 14 percent market share of the heating market. Electricity and gas transmission and distribution networks are regulated in German Energy Law but no such regulation exists for district heating; instead general rules of German competition law apply. Third party access to district-heating distribution grids is allowed but, at present, this has not been done to in any major extent.  The German government has put an emphasis on Combined Heat and Power and District Heating and Cooling as solutions for meeting environmental targets. A Combined Heat and Power Act (KWKModG) is planned to contribute to the objective of reducing annual carbon dioxide emissions. The Act also serves as the implementation of Directive 2004/8/EC on the promotion of cogeneration based on a useful heat demand in the internal energy market (Cogeneration Directive). The system shares some similarities with the Act on Granting Priority to Renewable Energy Sources (EEG) (\cite{ECOHEAT4EU}). \par

In Germany, the W{\"a}rmenetzsysteme 4.0 initiative has provided 100 million euro for funding feasibility studies and pilot projects related to low temperature heat recovery 4GDH. \par

\subsubsection{United Kingdom}
A key initiative in England and Wales is the Heat Network Delivery Unit (HNDU), which was formed within the Department for Energy and Climate Change (DECC) in summer 2013 (\cite{sneum2018policy}) to help meet 2050 targets. The Carbon Reduction Commitment (CRC) is a mandatory scheme that aims to improve energy efficiency and reduce $CO_{2}$. Eligible organisations must buy allowances to cover their reported emissions; hence, green DH investments have benefited from the scheme. Another mechanism with which to meet the 2050 goal is national and regional planning for infrastructure. The plans set the policy framework for infrastructure decisions (including heating infrastructure) through Planning Policy Statements (PPS). District heating plays a part in this scheme. These statements are combined with the Climate Change Levy (CCL) tax which taxes supplies of energy within the non-domestic sector including industry, commerce and the public sector. In the CCL, there are exemptions for fuel inputs and energy outputs from CHPs. \par 

An Enhanced Capital Allowances (ECA) scheme provides businesses with enhanced tax relief for investments in equipment that meet published energy-saving criteria. Eligibility for an ECA is a fiscal benefit available to new CHP schemes certified under the CHPQA programme. Since the most capital intensive element of distribution networks (the pipes) are not included, the scheme has limited overall effect in this sector. A number of funding programmes have been carried out in UK. These include both grant capital support and assistance with pre-investment activities. One example is the Community Energy programme that aims to deliver new community heating schemes and refurbish old ones, thus reducing carbon emissions, alleviating fuel poverty and reducing frontline energy costs. Another similar scheme was provided through the Homes and Communities Agency (\cite{aronsson2011existing}). \par

The government is progressing policy incentives that will reduce the heat demand of the existing building stock while promoting the uptake of renewable heating technologies. The recently deployed Green Deal is expected to remove the barrier of initial costs for energy efficiency improvements while the Renewable Heat Incentive (RHI) attempts to support market roll out of renewable heat technologies. However, the success of these policy initiatives is uncertain and the impacts on technology deployment are yet to be identified (\cite{ziemele2016effect}). The need for district energy market regulation has been discussed but, to date, there is no regulation of the heat market (\cite{sneum2018policy}). Taking the interest of low temperature district heating investments into account, it is notable that no regulation explicitly supporting such investments exists in the UK (\cite{reuseheat_2_1}). \par

\subsubsection{France} 
A large proportion of the supply of primary energy in France comes from nuclear power (\cite{percebois2012quel}). Use of renewables such as hydropower, wind and solar is growing, however. France is increasingly focusing on energy savings and the reduction of waste. The Grenelle laws focus on decreasing energy consumption in buildings. However, the use of central heating (\cite{belaid2016understanding}) makes the incentives blunt. France also offers tax credits for energy-efficient goods and zero-rate loans for energy efficiency renovations up to EUR 30,000 per dwelling.  The government has implemented energy performance contracts to be set up between owners and operators, establishing an energy efficiency target for a building.  Such contracts are increasingly considered by local authorities (\cite{eshien2013contrats}). 

There are two direct support measures for DH in France. The first one is reduced VAT (from 20 to 5.5 percent) for thermal heat delivery through DH systems. France also has a heat fund for heat from renewable sources (biomass, geothermal, biogas, PV). The idea is to provide incentives for increased shares of renewables in the fuel mix. Taking low temperature district heating investments into account, it is noted that no regulation explicitly supporting such investments exists in France (\cite{reuseheat_2_1}).

\subsection{Lessons learned}
District heating is widely acknowledged for its contribution to 2050 targets. Price regulation of heat has been implemented (Denmark) or discussed (Sweden) in two markets with explicit DH legislation. All five countries have put taxation in place to directly or indirectly support DH.  In two of the five countries, there is explicit support for low temperature heat recovery (Denmark and Germany). These conclusions are summarised in table~\ref{table:regulations} below. \par

\begin{table}[]
\fontsize{6}{6}
\selectfont
\label{table:regulations}
\begin{tabular}{|l|l|l|l|l|l|}
\hline
                                         & \multicolumn{2}{c}{Explicit DH legislation}    & \multicolumn{3}{|c|}{Energy and competition legislation}                \\
\hline
Policy type                              & Denmark         & Sweden                         & Germany                         & France                       & UK                  \\
\hline
2050 targets                             & -               & -                               & -                               & -                            & -                   \\
                                         &                 &                                 &                                 &                              &                     \\  
\hline
Price regulation                         & -               & discussed                       & NA                              & NA                           & NA                  \\
                                         &                 &                                 &                                 &                              &                     \\
\hline
Tax incentives                           & taxation makes  & tax on carbon                   & Electricity from CHP has        & Reduced VAT on               & climate change      \\
                                         & DHC competitive &                                 & low primary energy factor       & DH deliveries                & levy                    \\
\hline
Incentives for low  & 2012-2018          & -               & 100 million euro                & -                               & -                   \\
temperature (4 \& 5G)                    &                 &                                 &                                 &                              &                     \\
\hline
\end{tabular}
\caption{Lessons learned from frameworks in five countries}
\end{table}

The conclusions above are confirmed by a study performed in 8 EU countries, in which interviews were carried out with DH stakeholders about low temperature district heating investments. A total of 76 respondents took part in the study and one major conclusion was that waste heat (both industrial and low temperature) is not explicitly mentioned in existing support schemes for DH. Since there are explicit incentives for different forms of renewables, a competition between DH with waste heat and DH with renewables emerges. This is an unfortunate situation that brings us back to the purpose of this paper, that is to discuss the role of policy in future proofing the heating and cooling sector in Europe, by allowing urban waste heat recovery. \par


\section{Methodology}
The results in this paper derive from a number of sources described below

\subsection{Stakeholder Interviews}
An aim of the ReUseHeat project is to gather the views of various stakeholders and potential stakeholders of urban waste heat recovery.  Interviews were carried out with five types of stakeholders in eight EU countries (Sweden, Germany, Denmark, France, Italy, Spain, Belgium and Romania) and the results in this paper derive partly from these.  The types of stakeholder interviewed are as follows:

\begin{itemize}
\item policy makers
\item investors
\item district heating companies
\item waste heat owners
\item customers
\end{itemize}

A total of 76 respondents were interviewed. \par

\subsection{Other sources}
Impressions and information from the following sources have also been used:

\begin{enumerate}
\item Panel discussion at a ReUseHeat/HeatRoadMap joint event in Brussels in February 2019.  Of particular interest was a panel discussion that included a representative from Belfius Bank, a bank owned entirely by the Belgian state, that has expressed an interest in investing in waste heat recovery.
\item A ReUseHeat Policy Workshop held in Brussels in October 2019 bringing together policy makers, academics, investors and consultants, aimed at encouraging waste heat recovery investment. The event consisted of three sessions: (i) designing a legal framework, (ii) creating a track record for waste heat recovery and (iii) promoting financial support and guarantees.
\item Discussions with a representative from Caixa Bank, a not-for-profit financial institution based in Valencia, Spain, which has expressed an interest in investing in waste heat recovery.
\item Information gathered from the 39th Euroheat and Power Congress, held in Nantes in May 2019.
\item Information gathered from the 2018 Global District Energy Days held in Helsinki in September 2018.
\item Discussions with consultants at Nordic Energy and IMCG, Sweden.
\end{enumerate}

\section{Results and Discussion}

\subsection{The role of urban waste heat recovery in reaching 2050 targets}
A number of `Roadmaps' have been designed defining options for achieving 2050 goals. An important example is the European Union Roadmap 2050 project which sets out pathways
to achieve an 80 percent reduction in greenhouse gas emissions by 2050 (\cite{Roadmap_2050}). The Heat Roadmap Europe project (a series of commission funded projects mapping relevant heat sources for district heating) sets out approaches to the decarbonisation of the energy system in Europe. The examples provided in section~\ref{section:literature_review} indicate that DH can support reaching these targets. This is further confirmed by other sources of information. In 2050, the cost of carbon will be higher than it is today, fossil fuels will not be available and the fuels that are currently used (such as waste and biofuels) will be increasingly scarce. It is then that urban waste heat recovery will become crucial and, possibly, a critical component for reaching the desired targets. \par

\subsection{Creating awareness}
A significant challenge for waste heat recovery is a low level of awareness about the potential opportunities that it presents, both for owners of excess heat and for those that may wish to exploit it. Well targeted government policy may be able to improve this situation. For historical reasons, the prevalence of district heating varies significantly around Europe and the wider world and this has an important impact on the level of awareness of the opportunities of waste heat recovery. Sweden is the country in the world with highest volumes of industrial waste heat recovered in district heating systems. Even so, the volumes recovered are modest in comparison to volumes available. There is awareness about the available potential and there is knowledge about barriers in the research community whereas the awareness about waste heat recovery appears to be lower amongst other stakeholder groups. \par

The UK government has attempted to gauge awareness of waste heat recovery among businesses. A report published in 2016 found that awareness of waste heat recovery in the UK is mixed \cite{Barriers_enablers}. Companies whose energy costs were relatively high tended to have significant interest in energy efficiency measures and most had a good awareness of waste heat recovery. Among companies with relatively low energy costs, most of which were smaller, awareness of waste heat recovery tended to be low. \par

In a study of Swedish industrial waste heat recovery from 1974 onwards, it was found that large collaborations tended to be more viable than small ones. Collaborations involving recovery of large volumes typically had a longer lifespan than those involving low volumes (\cite{lygnerud2017risk}). It seems likely that the nature of low temperature heat recovery means that businesses are less likely to consider the idea without being prompt. It is perhaps intuitive that high temperature waste heat from, for example, heavy industry may have some value and that there ought to be some way of using it to increase the overall efficiency. The operator of a waste water treatment plant, on the other hand, is unlikely to be intuitively aware that their waste heat is worth anything at all. Increasing awareness for lower temperature heat sources may therefore be a bigger challenge. Despite this, one study on low temperature district heating business model development (\cite{lygnerud2018challenges}) shows that it is often the owner of the low temperature waste heat that approaches the district heating company asking for their support to make use of the waste heat generated. Again, in conclusion, urban waste heat recovery solution is seen as a future proof solution and to achieve a greater uptake, awareness of its potential is imperative. \par

\subsection{Capacity building}
Low temperature waste heat recovery is in its infancy as a technology and its prevalence is low. Currently, there are a number of barriers and government intervention may be required to break them down. Whilst few measures are in place targeting waste heat recovery specifically, there are proven government measures that can help in the context of low temperature district heating. Examples are direct incentives such as tax breaks and investment subsidies. Other measures directly targeting urban waste heat sources are either voluntary, such as pilot projects, or mandatory, such as obligations to make use of urban waste heat whenever generated. Implementation of urban waste heat recovery is, at present, predominantly undertaken in the form of pilot studies. In an ongoing project for the International Energy Association, more than 150 low temperature heat recovery pilot sites have been identified. The advantage of pilot sites is that they provide data on operations and risk which act as important inputs for potential investors. Formal assessment of the impact of a measure is crucially important to determining its effectiveness. Ideally, impact assessments should be performed before the measures are put in place and followed up once the policy is implemented. This can be done with mathematical models alongside risk assessment such as sensitivity analysis and Scenario Analysis. An important consideration is that different measures may interact with each other and therefore should be treated as part of a tree of decisions, rather than one stand alone decision. \par

The European Union's `Better Regulation' initiative performs impact assessments on proposed laws that are expected to have economic, social or environmental impacts (\cite{Better_regulation}). In addition, consultations with stakeholders and citizens are recommended. The EU commission also looks for areas of improvement in existing law. Key Performance Indicators (KPIs) are an important aspect of impact assessments, helping to quantify the performance of each measure. Often, multiple KPIs create trade offs. Perhaps the most prominent of these concerns a trade off between reducing greenhouse gas emissions and the overall cost of a project since a more `green' technology is often more expensive. A crucial target for government policy should therefore be to reverse this situation such that businesses are motivated to pursue green options. \par

The European Commission is directly funding two large demonstration projects on low temperature heat recovery, in a total of 12 sites across Europe (for more information please consult the Reuseheat and Rewardheat projects). Limits on greenhouse gas emissions by regulation would benefit urban waste heat recovery technology. Such regulation can be directly linked to taxes aimed at the phasing out of fossil fuels. Examples of both exist, as discussed in the policy overview above (the EU-ETS system) along with energy carbon taxes. In such a context, it is important to note that there may be social aspects to consider when making planning decisions. For example, the cost of decarbonisation should not disproportionately be borne by the poorest in society. In some cases, governments may see measures as opportunities to achieve social aims by, for example, building infrastructure to provide cheaper heat, paid for with taxation (a dominant tradition in both Sweden and Denmark). To secure desirable and long term development in the DH industry, support for capacity building in urban waste heat recovery is needed. \par

\subsection{What the investor wants}
A question often asked is whether consumer preference can be changed by appeal to public good, in this case represented by low carbon solutions. Put simply, are consumers willing to accept higher prices for types of energy with a lower carbon footprint? Classical economic theory base would say ``no'', because this would be a distortion of the market. A ``yes'' would imply that there are issues other than price which affect decisions. In fact, there is some recent evidence that public good arguments can change behaviour such as the successful recent campaigns to curtail the use of plastic bags. Such campaigns, however, may not be enough to solve the problems as it may be hard to lay the moral issues at the consumers' door. \par

In general, consumers tend to value causes in which they can `see' the impacts. The success of plastic bag campaigns has been the emotional reaction to stories of plastic pollution in the oceans, for example. In the context of waste heat recovery, it may be beneficial to emphasise the public good in a local context. The idea that locally sourced heat that would otherwise `go to waste' can be recovered and used instead of non renewables may be an appealing one. In some areas, this may create a demand for more expensive, but greener, heat. To reach 2050 goals, the nature of corporate governance must change from profit maximisation to climate optimisation. An obvious approach, introducing measures along these lines, is to `monetise' greenhouse gas emissions. The EU's carbon trading legislation, for example, places a monetary value on emissions, which can be traded at a cost, creating a financial incentive for reducing emissions. \par

There have been some attempts to prioritise public good in investment decisions. For example, the UK Public Services (Social Value) Act 2012, which applies only to government contracts, states that decision makers ``should be taking a value for money approach - not lowest cost'' and should therefore take into consideration any (positive or negative) externalities for the local area (\cite{Public_services}). At present, the act does not specify a value for the reduction of greenhouse emissions but this may be a valuable addition. \par

The European Union has also pursued `socially responsible' policy on procurement. In 2011, its `Buying Social' publication set out an agenda for taking social elements into account when considering procurement bids (\cite{european2010buying}). More recently, the EU have adopted reformed procurement rules that allow public authorities to take social elements into account when making procurement decisions. Guidance has been published, in the form of a ``Buying Green Handbook'', on how Green Public Procurement (GPP) can be implemented (\cite{official2011buying}). Although green considerations are, at present, voluntary, the EC aims to achieve a critical mass of demand for sustainable goods and services. The above legislation is not yet linked up to corporate governance, and to do so would require a heavy shift in stance. Such requirements are a far easier sell for government contracts than for private enterprise in general. However, it is easy to see such rules applying to carbon emissions, and legal requirements for a net zero-carbon economy by 2050 may make such legislation necessary. \par

In terms of policy, the most common approach is likely to be that which seeks to align the incentives of businesses and the public good. Taxes on carbon emissions, for example, attempt to make it financially advantageous for businesses to seek future proofed, low carbon alternatives. Policies that help reduce risk on environmentally friendly solutions should also be considered. For waste heat recovery, policy should be aimed at making it more financially viable than non-renewable alternatives. \par

Some forward looking financial institutions and companies have seen opportunities in investing in long term, low carbon, community based schemes.  One example is the collaboration between energy company Engie (France) and Axium that secured a 50 year Comprehensive Energy Management Contract with Ohio State University in the US in 2017. Another example is the collaboration between energy company VEKS (Denmark)and CP Kelco, a US-owned company that produces pectin, a natural starch. The company is located some 40 kilometres from Copenhagen and produces a large amount of excess heat. CP Kelco and VEKS have different business models. CP Kelco is an entirely commercial company and this means that, as a starting point, a large focus on investments with a short payback period is required. In contrast, VEKS, as an operator of infrastructure in the form of district heating systems, has a longer time horizon for its investments; VEKS works with more patient capital. An agreement was able to be reached due to clear system boundaries and a contract detailing payment schedules. CP Kelco was responsible for the investment and the design of the technical installations up to the "connection point" with VEKS. VEKS was responsible for the investments and the design of the technical installations from the "connection point" with CP Kelco and to the existing district heating network (meaning a larger investment for VEKS since about 150 meters of underground district heating piping was needed). In terms of payback, two time periods were established. In the first period of operation (3-4 years), CP Kelco received their payback and, in the second (some 4 years), the investment of VEKS was paid back.  \par

It is useful to think of DHC funding as a special type of infrastructure spending. This has the advantage of making it part of the wider and current discussion, both nationally and internationally, about the need to renew infrastructure and its role as one particular mechanism to stimulate economies. Wider energy and transport infrastructure are similar examples. The common features are that a mixture of public and private funding is often preferred with the notion that companies are reluctant to invest unless initial investment, or at least a high proportion of it, is provided by the public sector. Lump sum grants, land grants, or grants tied to project milestones are more closely related to direct financing instruments; they reduce the need for privately-sourced capital expenditures for the project and can also reduce initial outlay. This has the effect of enhancing returns to investors and can also enhance creditworthiness and the viability of the financing structure. \par

A generic term describing the area is ``project risk capital'' or ``risk capital funding schemes''. Related to this is the growth of special funds such as green funds, often managed at the city level (see \cite{MEEF,Goth_green_bonds} for example). A summary is that with (i) project, (ii) fund and (iii) institutional investors, there is a double effort: the fund needs to attract institutional investment and the project sponsors need to attract resources from the fund. The methodology also includes ``structuring'': matching the type of funding to the type of asset; in the case of DHC, this means long term funding. \par

In view of the urgency of the 2050 objectives, the role of infrastructure and social value objectives (the latter being very relevant because many DHC projects are private-public partnerships), we are likely to see an increase in the requirement to cut carbon in all DHC contracts, with associated incentives. This is a win-win: increased incentives will improve the profitability of the relevant contracts for the private sector while, at the same time, directing investment into carbon-free technologies. An example would be tax incentives or cheap loans for domestic or community heat pumps. There are also important links between national and local schemes. For example, local heat storage can be used to store surplus renewable energy from the national grid such as from wind. \par

A common theme coming from interactions with investors expressing an interest in waste heat recovery is that, although there is a desire to invest and that money is available to do so, project promoters do not adequately assess the investment risk. In short, investors need to be more reassured that there is a high probability of receiving their money back than at present. \par

Stakeholders in district energy need to manage different objectives, for example medium term lending horizons versus long term infrastructure investment. A conventional district heating network has a technical lifetime of approximately 40 years and the low temperature heat recovery investments have lifespans that are approximately half. The investors who are willing to lock in funds for decades are rare but as seen above there are ways to come around this parameter. \par

Investors generally look for mature technologies with low degrees of legal, technological and economic risk. The track record, experience and financial strength of project partners is a key aspect in the decision to invest along with predictability and stability of cash flows and stability in the political framework. Talking to investors on immature district heating market of the UK it was identified that even conventional district energy technology was perceived as new and risky. There is a knowledge gap that needs to be bridged between district heating practitioners  and investors interested in green. This is the case both for conventional district heating ventures and for the newer, low temperature solutions. \par

The concept of bankability, that is the extent to which a project is attractive to investors, must be part of a wider discussion of the risks of DHC projects. The risk analyses presented in business cases for funding should be close to those carried out by the potential funder. Stakeholders, however, need to manage different objectives, for example medium term lending horizons versus long term infrastructure investment. Technical risks arise from the novelty of projects, which feed off inventive sources of waste heat, and a lack of experience, which, itself, is a major source of risk. There is experience of waste heat recovery from data centres and sewers but not of, say, hidden underground rivers (\cite{Lost_rivers}). \par

Even if the styles of risk analysis and the KPIs are shared between the funder and funded there remains some macro issues to be taken into account, most importantly zero carbon objectives and road maps. On the positive side, for some of these, we are all in the same boat, or in this case planet. The vehicle for trying to handle these larger scale issues is scenario analysis (SA) which has grown rapidly in the last few years particularly in the area of climate and energy. The Intergovernmental Panel on Climate Change's IS92 comprised six scenarios which were followed by the “Special Report on Emissions” (SRES) and the ``Representative Concentrated Pathways'' (RCPs). \par

It is becoming more common for mainstream risk methods to be supplemented with SA. This could be referred to as global sensitivity analysis as opposed to the simulation based local sensitivity analysis carried out on the spreadsheets of CAPEX and OPEX. Financial Institutions are familiar with this under the heading of the stress testing required by regulators. SA can be seen to fill, however roughly, the gap between local sensitivity, amenable to various statistical methods, and a less easily definable uncertainty about the future. All actual risk is predicted on particular scenarios. Some scenarios refer to future policies, such as tax incentives, whilst others may be contingent on events such as a sudden and unforeseen increase in electricity prices. \par

Actual investment needs to checked against an agreed tapestry of scenarios. The agreement is that the investment needs to be robust, not just to local variation but against agreed scenarios, and the risk allocation made accordingly. The technical side of SA is a major research area but can at least be divided into various headings such as scenario design and scenario control and it is generally agreed to be a useful creative tool (\cite{wheatcroft2019scenario}). \par

DHC itself is more closely related to climate scenarios, given the ``climate emergency''. This is likely to play a major role in the new accelerated agenda and the shared responsibility of cities and their funders ought to be a driver for investment, to add to the special incentives and stricter laws. \par

The ReUseHeat project has favoured the use of modelling as a foundation for contracts and demand forecasting.  This extends to risk.  Cities need to improve understanding of the risk of a project, particularly to facilitate the route from feasibility studies to a business case.  The expertise that cities have may be less than that of financial institution but increased harmony between the risk methods would be valuable for bankability.  The contractors are experts at physical risk whereas the financial institutions are able to tier the financial risks and these are not the same thing. \par

Given the increasing insecurity arising with climate change, it is noticeable how the use of scenarios has gathered strength.  The methodology of how to design and make use of scenarios is very much less developed than risk theory.  Despite this, energy companies and governments are working with them in order to supplement more formal risk analysis.  The kinds of local sensitivity analysis which are commonplace in dealing with risk based spreadsheets needs to be extended to robustness with respect to major scenarios. We have already seen this with wind power where rapid development is faster than predicted by forecasts.  Storage, both of electricity and heat is developing similarly quickly.  We feel that, rather than scenarios describing events which instil anxiety, it is important that they be presented as commercial opportunities.  Whilst waste heat recovery may be a more expensive option at present, future policies may reverse this situation and investing now may give companies a competitive advantage in the future.  There may also be other opportunities such as selling heat storage to the national grid to observe peak outputs of renewables.  Every data centre, metro system and canal is an opportunity to use heat pumps to provide heat.  \par

\subsection{Legal Framework}
The lack of a systematic approach to regulation is a major, perhaps the biggest, impediment to investment in DHC, and to urban waste heat recovery in particular. Indeed, regulatory risk (also referred to as political risk) is usually included in risk analysis. A systematic approach to DHC should be part of a systematic approach to energy in general and the fact that low temperature DH consists of small local schemes does not aid integration. Local advantages may ignore negative externalities and national initiatives may ignore local DH, even to the extent that incentivising electricity may ignore heat and cooling, giving incentives for CHPs but not for heat pumps and waste heat recovery systems. It is very likely that this will be rectified when the importance of local DH to international roadmap objectives
is realised. \par

An example of where a broad brush approach at the national level ignores DHC is in the Third Party Access (TPA) directives. Another lack of clarity is that diseconomies of scale may make it costly to produce heat at a local scale, rather than on a larger scale. Thus, incentives may apply when a heat pump is domestic but may not be valid for larger heat pumps at a community scale, which may be more efficient. In many cases, excess heat is not properly recognised as a heat source and there are no incentives at all. This ought to change as its contribution to $CO_{2}$ reduction is properly recognised. DH is, at present, routinely ignored in the climate debate. This is surprising because the DHC system can use sources such as geothermal, biomass and solar energy. At a technical level, energy efficiency indicators for buildings should take into account total energy consumption rather than just primary fossil fuel. \par

\section{Conclusions and Policy Implications}
We make three policy recommendations and these are described below:

\begin{enumerate}

\item \emph{Create awareness and a standardised legal framework so that DH can support the attainment of the 2050 goals.} \\
There is limited knowledge in Europe of the opportunities that waste heat recovery offers. Understanding of what the heating market might look like in 2050 is also limited. In 2050, there will almost certainly be no fossil fuels, alternative fuels for heat generation will be limited (waste and biofuels) and the cost of carbon will reflect the cost to the environment. Awareness about the future and options for heating needs to be created. With greater awareness, urban waste heat recovery will become a natural step towards future proofing the district energy sector. \par

The low level of awareness of waste heat recovery is reflected by the absence of a legal framework. This makes contracting and investments complicated and time consuming.  The inclusion of waste heat (including urban waste heat) into the decarbonisation strategy of the EU would reduce the risk of competition between waste heat recovery and renewables. \par

\item \emph{Strengthen capability for undertaking urban waste heat recovery.}
\\
We recommend that the implementation of pilot projects should be heavily supported by policy makers. Such projects allow for the technical and economic feasibility of waste heat recovery projects to be demonstrated in terms of how closely they replicate the planning stage. Strengthened capabilities in urban waste heat recovery generate both technical and non-technical knowledge which can be Incorporated into future projects, reducing the risk. The involvement of utility companies or other long term investors can be an advantage in terms of risk reduction. It is worth noting that pilot projects do not need to receive funding from public entities and may, instead, be funded by private companies if they are motivated to do so. \par

\item \emph{Bridge the gap between urban waste heat recovery projects and investors.}
\\
There is a desire from decision makers in Europe to reach 2050 targets. This means that investors are typically interested in green projects. Urban waste heat recovery projects have a significant green value. Hence, from an economic point of view, there appears to be both supply of demand for green investments. Within the ReUseHeat projects, it has been found that low maturity of urban waste heat recovery investments requires support from the district heating community to make investors understand the business case, risk and the value of green. A possible approach to bridging the gap between urban waste heat recovery projects and investors was identified at the ReUseHeat policy workshop held in October 2019 in Brussels. A pilot credit facility is recommended, that covers some portion of the risk for waste heat recovery projects. The facility would have a total size of 20-30 million euro with a maximum of 1-2 million euro per project and would help bridge the knowledge gap between urban waste heat recovery projects and potential investors. \par

\end{enumerate}

\section*{Acknowledgements}
This work was funded by the ReUseHeat project, grant number 767429.

\bibliographystyle{abbrv}
\bibliography{paper}

\end{document}